\documentstyle [11pt,epsf]{article}  
\title{${\rm Schr\ddot odinger\/}$ Equation and Phase Space \\
          in Quantum Mechanics \\[1.0cm] }
\author{ \it{ Kiyoung Kim} \\
        \it{Department of Physics, University of Utah, SLC, UT 84112}}%
\begin{document}
\pagenumbering{arabic}
%
%\vspace{10 pt}
%\date{April 14, 1997}
\maketitle
\vspace{20 pt}
\begin{abstract}
\noindent 
Using classical statistics,  ${\rm Schr{\ddot o}dinger\/}$ equation 
in quantum mechanics is derived from {\it  complex space\/} model\cite{kim}. 
Phase-space probability amplitude, that can be defined on classical point of 
view, has connections to probability amplitude in  internal space and to wave 
function in quantum mechanics.   In addition, 
the physical entity of wave function in quantum mechanics is confirmed 
once again.      

\end{abstract}
\vspace{20 pt}
\def\vac.{{vacuum}}
\def\ivac.{{\it vacuum\/}}
\def\SP{{special theory of relativity }}
\def\Sp{{Special theory of relativity }}
\def\QM{{quantum mechanics }}
\def\Qm{{Quantum mechanics }} 
\def\#{{\makebox[0.9mm]{}}}
\section{Introduction}
\paragraph{}  
Since E. Wigner and L. Szilard proposed the phase-space distribution
\cite{wig}\cite{hil}  to study quantum corrections to the classical statistical
 mechanics in 1932, the distribution({\it Wigner Distribution\/}) has been 
studied in quantum optics,  quantum chemistry, and {\it etc.\/} In 1947, Moyal, 
J. E.\cite{moyal} attempted to interpret \QM  as a statistical theory,  and 
some other people have tried it with the distribution or using some other way. 
Especially, L.S.F. Olavo made a bridge from classical statatistics to quantum 
theory using the transformation, {\it Wigner-Moyal\/} Infinitesimal 
transformation\footnote{the name was cited by L.S.F. Olavo.} and Liouville's 
theorem in classical phase space\cite{ola1}\cite{ola2}.  
It was a brilliant insight in physics.    
But physical meaning of the expressions, especially ontological point of view, 
has not been interpreted yet. 
\paragraph{}  
In the paper, {\it The Wave Function in Quantum Mechanics\/}\cite{kim},  we 
supposed that  physical space consists of {\it 4-dimensional complex space\/} 
and that there are interactions among real physical object 
and \ivac. {\it particles\/}(negative energy mass electrons).  Also we assumed 
that electromagnetic energy is propagated through \vac.-particle-string  
oscillations. 
\par
\medskip 
In those assumptions, words, {\it intangible\/} and {\it imaginary world\/} 
were used because we can not prove the exsistence directly through experiments 
and also because we have not recognized the interactions with \ivac.   
particles.  Even in the definition of word, {\it phenomena\/}, we have not 
included or connected to those particles(not antiparticles). 
Being considered that the {\it complex space\/} was supposed not for 
a mathematical tool, but for a genuine physical space in ontology, 
those intangible particles in imaginary world should be, ultimately, 
included in phenomena through a  physical conception. 
\paragraph{}   
In former paper\cite{kim} ; first, Plank's constant(h) was investigated in 
special theory of relativity,  then the physical meaning of {\it photon\/} 
was interpreted.  
Finally, the physical entity of wave function in \QM  was searched through 
new interpretation of special theory of relativity.   Furthermore, we could 
have a clue which can explain {\it the second law of thermodynamics\/} if we 
consider that the interaction with \vac.  particles is unavoidable, thus    
a physical system cannot be isolated  from \vac. particles.   However, 
one of conclusions is following : When physical object has kinetic energy, 
it is wrapped, in the respect of phenomena,  with \vac. particles
(\vac. electrons in the model),  those of which are interacting with 
the physical object; the effect of interaction is following the physical 
object in a wave form through \vac.-particle-string oscillations
({\it i.e.\/} transverse mode)  with velocity $v_g$, that is the same as that 
the physical object has.  In addition, those \vac. particles, in fact, 
carry the physical momentum p, that is the momentum of the physical object  
conventionally  and also they have kinetic energy, pc.  
\par
\medskip
Even though the physical entity of wave function in \QM  was interpreted as a 
representation of  interaction with \vac. particles  wrapping and 
following the physical object in {\it phenomenological view\/},  actually 
the interaction propagates in an imaginary subspace.     
\paragraph{}     
In Section 2,  ${\rm Schr{\ddot o}dinger\/}$ equation is derived using 
classical statistics or fundamental statistical concept.  In the process, 
 we can find that phase space representation is also possible, definitely, 
thus  in the formalism, it cannot be discriminated in the name of uncertainty 
principle. In addition, the physical meaning of {\it Wigner-Moyal\/} 
Infinitesimal transformation or characteristic function\cite{ola2}  is 
interpreted with classical statistics and  complex space model.   
Finally, Summary is followed.

\section{Classical Statistics}
\subsection{classical phase-space and ensemble} 
\paragraph{} 
In classical phase space, the physical meaning of an {\it ensemble\/} is a  
geometrical point set(mental copies of a system), each point of which is 
corresponded to a same, can not be distinguished, macroscopic phenomena. 
Usually, we define the  ensemble for a many particles system, and then use 
ensemble average for macroscopic phenomena which represents the system. Also 
we can define probability density with the ensemble to characterize local 
properties of the system.  
\paragraph{} 
For example, to describe non-relativistic, one dimensional, and one particle 
Newtonian system with a hamiltonian, $H(P,X)$,   a classical phase-space can 
be chosen for a deterministic representation  even though it is skeptical 
if the representation is complete or not.  
Once we accept the interaction with \vac. particles, the one particle 
description in the phase-space is not complete,  because  non-relativistic 
Newtonian system is considered as a limiting case when $v/c \#\ll\#1$ in which 
light velocity(c) never be infinity.  By the way, we can understand how 
non-relativistic Newtonian system is related with interactions with  \vac.  
particles  from the relation of universal constants, $c$ and $h$  
as\cite{kim}   
\begin{equation}
{h \over c} \#=\# {p_\lambda \lambda \over c}
            \#=\# 2 \pi^2 m_e \# \left( {A^2 \over d} \right), 
\label{UNIV}
\end{equation}
where $A$ is the amplitude for each \vac.-particle-string oscillation,  
$d$ is the equilibrium distance among \vac. particles, 
and $p_\lambda$ is the momentum of one wave length in a \vac.-particle-string 
oscillation.    
If Plank's constant, $h \rightarrow 0$ or light velocity, 
$c \rightarrow \infty$ in eqn(\ref{UNIV}), the amplitude, $A$ on RHS should be 
zero, thus  the interaction effect  disappear. But in reality, the effect from 
the interaction with \vac. particles cannot be zero although it can be ignored 
in an approximation if it is so small relatively.  
\subsection{One particle system(1-dim.,time independent Hamiltonian)}  
Alternatively, we can choose a phase-space with  \vac. particles and 
use classical statistics as a non-deterministic dynamics.        
In complex space, each real coordinates($\vec x$) has a 3-dimensional 
imaginary subspace and  interactions of the real physical object with \vac. 
particles in the subspace follow the physical object in wave forms, that is, 
infinity number of \vac.-particle-string oscillations.  
Hence, for one dimensional one particle(real physical object) motion,  
we can imagine one string-wave corresponed to each real coordinate 
with momentum, $p$.   Here, {\it superposition\/} principle is presupposed  
to describe the string-wave as 
\begin{equation} 
\phi(x,p;t) \#=\# {1\over \sqrt{2 \pi \hbar}} 
                \int \xi(x,x';t) \# e^{-{i\over \hbar} p x'} \# dx', 
\label{PROA}
\end{equation} 
where $x'$ is in the direction of  which  string-waves  propagate in 
the 3-dimensional imaginary subspace.  we might wonder that 
${p x'\over \hbar}$ in eqn.(\ref{PROA}) is not ${p_\lambda x'\over \hbar}$ 
since wave number, $k = p_\lambda / \hbar$  in eqn(\ref{UNIV}).  
Because the integration of $x'$ in eqn.(\ref{PROA}) is from $-\infty$ 
to $+\infty$, with  a scale transformation of $x'$ \# $p_\lambda$ can be 
substituted with $ p$, that is a possible net momentum in the string wave 
at coordinate $x$.  
\par 
\medskip 
$\phi(x,p;t)$ is characterized by $\xi(x,x';t)$, that is amplitude related 
with the imaginary coordinate $x'$ in internal complex space$(x,\# ix')$,  
to describe the string-wave corresponded to real coordinate, $x$.  
The complex function, $\xi(x,x';t)$ is not defined in the internal complex 
space since $\xi(x,x';t)$ came from a functional relation between 
the imaginary coordinate, $x'$ in the internal complex space and the momentum, 
$p$ in the phase space at real coordinate, $x$. 
Hence, {\it analytic\/} condition  is not necessary to the complex function, 
$\xi(x,x';t)$. Let us define {\it internal space\/},$(x,x')$ for the function, 
$\xi(x,x';t)$. 
\par
\medskip   
Since electromagnetic waves  was supposed as \vac.-particle-strings 
oscillation\cite{kim},  it is natural to assume that $\phi^{\dagger}(x,p;t) 
\phi(x,p;t)$ is proportional to energy density or probability density 
without a loss of generality.  Thus, the name, probability amplitude in  phase 
space is appropriate.  Let us call $\phi(x,p ;t)$ and $\xi(x,x';t)$ 
{\it probability amplitude\/} and {\it characteristic amplitude\/}, 
respectively.\cite{ola1}  In addition, $\phi(x,p;t)$ is supposed 
as a continuous function of $x$ and $p$.     
\par 
\medskip 
Now we have infinite number of string-waves, each of which is corresponded to 
a real coordinate.  Therefore, an {\it ensemble\/} in the phase-space is made 
up with infinite number of string-waves with a momentum $p$ distribution at 
coordinate  $x$.   Hence,  probability density in  phase-space can be defined 
as  
\begin{equation} 
F(x, p; t) \#=\# \phi^{\dagger}(x,p;t) \# \phi(x,p;t),
\label{PROP} 
\end{equation}     
in which $(x,p)$ is an abbreviation of continuous and infinite dimensions.  
And  macroscopic phenomena,  which is represented with coordinate and momentum 
of the real physical object,  have following relations, 
\begin{equation}
{\langle p(t) \rangle} \#=\# \int p(t)\# F(x,p; t) \#\# dx dp  
\label{AVP}
\end{equation}
and 
\begin{equation}
{\langle x(t) \rangle} \#=\# \int x(t)\# F(x,p; t) \#\# dx dp,    
\label{AVX}
\end{equation}
as  {\it ensemble\/} average.  In a non-relativistic case, we can expect 
that ${\langle p(t) \rangle} \sim  P(t)$  and ${\langle x(t) \rangle} 
\sim X(t)$ as mentioned before.  In addition, If we remember that total 
momentum of string waves is equal to the momentum  of real physical object 
in relativistic or non-relativistic case\cite{kim},  eqn(\ref{AVP}) is 
self-consistent.    
\subsection{Wigner-Moyal Infinitesimal Transformation}
For any non-relativistic system,    
the definition of Wigner-Moyal Infinitesimal Transformation\cite{ola1} is  
\begin{equation}
\rho\left(x+{\delta x \over 2}, x-{\delta x \over 2}; t \right) \#=\# 
    \int F(x,p;t) \exp \left(i {p\delta x \over \hbar} \right) \# dp. 
\label{WM}
\end{equation} 
This definition is consistent to  Wigner Distribution\cite{wig} 
except its infinitesimal nature.  From eqn.(\ref{PROA}) and eqn.(\ref{PROP}),  
the Infinitesimal Transformation in eqn.(\ref{WM}) can be expressed with  
a couple of characteristic  amplitudes, that is 
\begin{eqnarray}
\makebox[1cm]{} \rho\left(x+{\delta x \over 2}, x-{\delta x \over 2}; 
                          t \right) 
   &=& \int \xi^*(x,x';t)\# \xi(x,x'+\delta x; t) dx' \nonumber \\
   &=& \int \xi^*(x,x';t)\# e^{{i\over \hbar} \delta x P_{op}} \# 
            \xi(x,x'; t) dx',   
\label{PAVE} 
\end{eqnarray} 
where operator, $P_{op} = - i \hbar {\partial \over \partial x'}$.  
\par 
\medskip
In the definition of probability amplitude, $\phi(x,p;t)$, in eqn.(\ref{PROA}),
 momentum $p$ and coordinate $x'$ are related with Fourier transformation.  
That means, if $\phi(x,p;t)$ is a probability amplitude in  phase space  
$(x,p)$, then $\xi(x,x;t)$ is also probability amplitude in internal space, 
$(x,\# x')$.  So, we can define the operator, $P_{op}$  
as in eqn.(\ref{PAVE}) to calculate $\langle p^m \rangle$, where $m$ is 
any integer in principle.  With the characteristic amplitude, that is 
probability amplitude now  in  internal space  $(x,\# x')$, the 
calculation of averages, $\langle x^n \rangle$, $\langle p^m \rangle$, and 
$\langle x^n p^m \rangle$ or $\langle p^m x^n \rangle$ is straightforward : 
\begin{equation} 
\langle x^n \rangle \#=\# \int \xi^*(x,x';t) \# x^n \#\xi(x,x';t) 
                           \#dx' dx, 
\makebox[2cm]{} 
\label{CALC1} 
\end{equation}
\begin{equation} 
\langle p^m \rangle \#=\# \int \xi^*(x,x';t)\# 
       \left(- i \hbar {\partial \over \partial x'}\right)^m \# 
       \xi(x,x';t) \#dx' dx, 
\makebox[4mm]{}
\label{CALC2} 
\end{equation}
and 
\begin{eqnarray} 
\langle x^n p^m \rangle \#&=&\#\langle  p^m x^n \rangle  \nonumber \\ 
       \#&=&\#  \int \xi^*(x,x';t)\# \left[ x^n  
           \left(- i \hbar {\partial \over \partial x'}\right)^m  \right] \# 
           \xi(x,x';t) \#dx' dx,  
\label{CALC3} 
\end{eqnarray}    
without any ordering problem.  
\par
In eqn(\ref{PAVE}) we can interpret the infinitesimal transformation as  
an  expectation value for  infinitesimal translation of $\xi(x,x';t)$ 
in the internal  space, $(x,x')$.  In addition, from eqn.(\ref{WM}) and 
eqn.(\ref{PAVE}), the infinitesimal translation is represented 
in the phase-space as a symmetry  regardless of the sign of $\delta x$,  
but actually  it is expressed  as a summation of the probability density with 
infinitesimal oscillations at coordinate, $x$.   
\subsection{Complex Function, $\xi(x,x';t)$  in Internal Space}
In former paper\cite{kim}, we assumed 7 propositions as the  starting point. 
One of them : Any change in imaginary world reflects to real world  and 
{\it vice versa\/}. This means {\it  symmetry\/} and {\it duality\/} of Nature. 
\par
If complex function $\xi(x,x';t)$, for example, can be defined in internal  
space $(x,\# x')$, the function, $\xi(x,x';t)$, must be symmetry 
under the variable exchange($x\#\leftrightarrow \#x'$).   
As a possible choice, we can define   
\begin{equation} 
\xi(x,x';t) \#\equiv\# \psi(x;t)\#\psi(x';t). 
\label{SEPA} 
\end{equation} 
It is a separation of variables  and also  satisfied with the symmetry.    
With the definition in eqn.(\ref{SEPA}) the averages, $\langle x^n p^m 
\rangle$ and $\langle p^m \rangle$ can be expressed as 
\begin{eqnarray} 
\langle p^m \rangle \#&=&\# \int \psi^*(x';t)\#\left(- i\hbar {\partial \over 
                 \partial x'}\right)^m \# \psi(x';t) \#dx',\makebox[3.6cm]{}   
\label{SAP} 
\end{eqnarray}  
and 
\begin{eqnarray} 
\langle x^n p^m \rangle \#&=&\# \int \psi^*(x;t)\#x^n\# \psi(x;t) \#dx \#\# 
                              \int \psi^*(x';t)
                \#\left(- i\hbar {\partial \over \partial x'}\right)^m \# 
                              \psi(x';t) \#dx'  
\label{SAXP} 
\end{eqnarray} 
with  $\int \psi^*(x;t) \psi(x;t) \#dx \#=\# 1$.  
\paragraph{}
In internal  space, there is no problem in operator ordering   
since momentum operator, ${\hat p} = -i \hbar {\partial \over \partial x'}$
is defined with  coordinate, $x'$.      
In addition, if this $2$-dimensional  formalism is brought to 
$1$-dimensional real space  formalism to use only real coordinate  $x$,  
we cannot find any problem as long as there is no operators {\it coupled\/}, 
such as $\hat x\#^n \hat p\#^m$, $\hat p\#^m \hat x\#^n $, and {\it etc.\/}  
\paragraph{} 
If a system is described with non-relativistic  $1$-dimensional  Newtonian 
kinetics for one particle, the energy of the particle,    
\begin{equation} 
E \#=\# {p^2 \over 2 m} \#+\# V(x), 
\label{HAMIL}
\end{equation}
where $m$ is mass, $V(x)$ is a potential function, of the particle.  
Using  eqn.(\ref{SAP}) the average energy,  
\begin{eqnarray} 
\langle E \rangle \#&=&\# \int \psi^*(x';t)\#\left({\hat p\#^2 \over 2 m}
          \right)\# \psi(x';t) \#dx' 
             \#+\# \int \psi^*(x;t)\# V(x) \# \psi(x;t) \#dx, \nonumber \\
        \#&=&\# \int  \psi^*(x;t)\# \left[{1\over 2m}
                \# \left( -i\hbar {\partial \over \partial x} \right)^2 
                \#+\# V(x)\right] \# \psi(x;t) \#dx,  
\label{SCH} 
\end{eqnarray}   
where $\hat p = -i\hbar {\partial \over \partial x}$.  
\par
\medskip
Now we can express eqn.(\ref{HAMIL}) with operators, 
$\hat p = -i\hbar {\partial \over \partial x}$, and $\hat x = x$.  That is  
$$ 
E \# \psi(x;t)\#=\# {\hat p\#^2 \over 2 m}\#\psi(x;t) \#+\# V(x)\#\psi(x;t),  
$$
This is nothing but ${\rm Schr\ddot odinger\/}$  equation.   
\paragraph{} 
In the process, we supposed that the complex function, $\xi(x,x')$ in internal 
space is separable as  $\psi(x)\psi(x')$.  
But we can choose it as like $\psi^*(x)\psi(x')$ since there is no change in 
the results.  However, we expect {\it analytic condition\/} when the function, 
$\psi(x)\psi(x')$ is mapped from internal space $(x,x')$ to inernal complex 
space $(x, ix')$. That means, complex function  $\psi(x)\psi(ix')$ should be 
analytic.   For instance,  the wave function of a free particle is  
$$ 
\psi(x) \sim e^{i K x}  \makebox[1cm]{} 
\makebox[5cm]{($K = {P\over \hbar}$, \#$P$ is momentum)} 
$$
Then $\xi(x,x') \sim  e^{i K x} \# e^{i K x'}$, and through the mapping, 
that is, 
$\xi(x,x')$ in $(x,x')\#\# \rightarrow \#\# \zeta(x,x')$ in  $(x,i x')$, 
\begin{eqnarray}
\zeta(x,x') &\sim& e^{i K x} \# e^{- K x'} \nonumber \\ 
            &=& e^{iK\#(x+ix')},           \nonumber 
\end{eqnarray}
that is analytic.  If we choose $\xi(x,x')$ as like $\psi(x)^*\#\psi(x')$, 
it corresponds a conventional change in mathematics , such like   
{\it right hand rule\/} or {\it left hand rule\/} in physics.

\section{Summary} 
${\rm Schr\ddot odinger\/}$ equation({\it time independent\/}) was derived 
with a classical statistics and complex space model\cite{kim}.     
Even though one dimesional one particle Newtonian system was used, 
the formalism is  general.  
Therefore, it is straightforward for 3-dimensional and multiparticle  system. 
\par \medskip  
We became to know that the probability amplitude in phase space is connected 
to the wave function in quantum mechanics through internal space$(x,x')$.  
Once again we confirmed the physical entity of wave function in quantum 
mechanics.
\paragraph{} 
It is about time to review those above results with specific cases in detail. 
Furthermore, once we accept complex space as physical space itself, it is not 
so hard to figure out the spin of electron in ontological point of view, that 
is 2-dimensional complex space rotation.  Also we, hopely, can understand more 
closely Pauli exclusion principle, Berry space, and their connection. 
 
%

%\newpage
%\input{figcap.tex}              
\end{document}